# Defect-Mediated Stability:
# An Effective Hydrodynamic Theory of Spatiotemporal Chaos

Carson C. Chow[*]

*Department of Astrophysical, Planetary, and Atmospheric Sciences*

*University of Colorado, Boulder, Colorado 80309*

Terence Hwa[**]

*School of Natural Sciences*

*Institute for Advanced Study, Princeton, New Jersey 08540*

(December 8, 1994)

## Abstract

Spatiotemporal chaos (STC) exhibited by the Kuramoto-Sivashinsky (KS) equation is investigated analytically and numerically. An effective stochastic equation belonging to the KPZ universality class is constructed by incorporating the chaotic dynamics of the small KS system in a coarse-graining procedure. The bare parameters of the effective theory are computed approximately. Stability of the system is shown to be mediated by space-time defects that are accompanied by stochasticity. The method of analysis and the mechanism of stability may be relevant to a class of STC problems.

# I. INTRODUCTION

A major goal in the study of spatiotemporal chaos (STC), the chaotic dynamics of a spatially extended deterministic system, is to obtain a statistical description in the limit of large length and long time scales (the hydrodynamic limit) [1,2]. One avenue towards this end is the use of effective stochastic equations. Once effective equations are established, standard methods of statistical field theory can be employed to described the behavior in the hydrodynamic limit [3]. Some outstanding issues include understanding when the stochastic descriptions are applicable and how hydrodynamic equations may be constructed.

Here, we construct an effective hydrodynamic theory for the Kuramoto-Sivashinsky (KS) equation in 1+1 dimensions [4–6],

$$\partial_t h = -\nu \partial_x^2 h - \kappa \partial_x^4 h + \frac{\lambda}{2}(\partial_x h)^2, \qquad (1.1)$$

which, due to its simplicity, has often been regarded as a paradigm for STC. The KS equation (with $\nu, \kappa > 0$) is believed to describe a variety of systems driven away from equilibrium. Among them are flame front propagation [4], the fluctuations of a fluid flowing down an inclined wall [5], and phase turbulence in convection cells [6].

The qualitative phenomenology of STC in the KS equation is rather simple: The system is linearly unstable to long-wavelength perturbations, and locally forms "cells" of a preferred size, given by the most unstable wavelength. It is however (exponentially) unlikely for these cells to be in complete registry in a large system [7] because they initially develop independently in different regions according to the linearized equation. Generically, the cells are locally compressed or stretched from the preferred size. Too much compression eventually leads to *cell annihilation* in the nonlinear regime; similarly too much stretching leads to *cell creation* [2]. These annihilation and creation events are "space-time defects" formed to reduce the local cell misalignment. However, the defects generally do not improve cell alignment at the global scale but serve to generate more defects. Thus the chaos is characterized by local cellular patterns interspersed with dislocations [8]. Figure 1 shows a spacetime profile of the chaotic state. Defect-mediated or phase turbulence is a term sometimes used to describe this situation [1].

Previous studies of the KS equation include aspects of the statistical properties [9–12], the spectrum of Lyapunov exponents [13], the attractor/bifurcation map for small systems [14,15], and the stability analysis of the stationary solutions [7,16]. Ref. [1] provides a recent review. It was conjectured by Yakhot [17] that in the hydrodynamic limit, the probability distribution of the KS equation may be mimicked by that of the noise-driven Burgers' equation [18],

$$\partial_t h = \nu_{\text{eff}} \partial_x^2 h + \frac{\lambda_{\text{eff}}}{2} (\partial_x h)^2 + \eta(x,t), \qquad (1.2)$$

where $\eta$ is an uncorrelated Gaussian noise, with second moment

$$\langle \eta(x,t) \eta(x',t') \rangle = D_{\text{eff}} \delta(x-x') \delta(t-t'). \qquad (1.3)$$

The effective parameters $\nu_{\text{eff}}$ and $D_{\text{eff}}$ are assumed to be positive constants, and $\lambda_{\text{eff}} = \lambda$. It is useful to think of $h$ as describing the height profile of a driven interface. In this context, Eq. (1.2) is also called the KPZ equation [19,20], whose solution's are well known in one dimension.

Yakhot's conjecture was motivated partly from symmetry considerations. The KPZ equation is the simplest equation which preserves two important symmetries of the KS equation, the translational symmetry $h \to h + \text{Const}$, and an infinitesimal Galilean symmetry $h \to h + \theta x$, $x \to x + \lambda_{\text{eff}} \theta t$, with $\lambda_{\text{eff}} = \lambda$. In addition, the apparent stability of the solution at large scales demands a *positive* surface tension, $\nu_{\text{eff}}$, while the chaotic degrees of freedom motivates a Langevin noise. This conjecture was found to be consistent with results of large scale numerical simulations of the KS equation [11,12]. However, an explicit demonstration of the connection between KS and KPZ has been lacking. In particular, the precise microscopic mechanism of generating stability ($\nu_{\text{eff}} > 0$) is not well understood.

In this paper, we show that the stability of the KS system at large scales is mediated by the space-time defects. We have dubbed this phenomenon "defect-mediated stability".

The chaotic nature of the defect dynamics then implies that stability and stochasticity are unavoidably interconnected. To obtain these conclusions, we constructed an effective hydrodynamic theory by explicitly coarse-graining the KS equation and utilizing the statistical properties of the chaotic dynamics in *small* KS systems (on the order of a few cell sizes). The method and the stability mechanism may be relevant to a class of STC problems.

We begin in Sec. II with a review of some basic properties of the KS equation. In Sec. III, we re-examine the perturbation theory about the stationary cellular solutions, where space-time defects are excluded by definition. We illustrate how the dynamics of a large system may be obtained from various properties of small systems, and show the lack of diffusive stability (surface tension) in the large scale dynamics to be the direct consequence of a conservation law and symmetries pertaining to the cellular state. In Sec. IV, we extend the analysis to the chaotic regime and show how a stabilizing effective surface tension $\nu_{\text{eff}} > 0$ is generated from the occurrence of space-time defects. We propose a coarse-graining procedure which allows us to construct the effective hydrodynamic equation from the chaotic dynamics of *small* KS systems subject to external forcings. Relevant statistical properties of the small system are obtained numerically in Sec. V. In Sec. VI, we analyze the results which lead to the KPZ equation as the effective stochastic equation in the hydrodynamic limit, and yield the approximate values of the bare KPZ parameters. A summary of the main ideas, the methods, the key assumptions, and the major results is given in Sec. VII, where we also discuss possible ramifications of our approach to other STC problems. A review of stochastic dynamics is given in the Appendix.

## II. BASIC PROPERTIES

We begin with a review of the linear stability analysis. For small fluctuations in $h$, the nonlinear term in Eq. (1.1) can be neglected. In this case, the linear equation exponentially amplifies all modes with wavenumber $k < k_1$, where $k_1 = \sqrt{\nu/\kappa}$. The fastest growing mode is $k_0 = \sqrt{\nu/2\kappa}$, corresponding to a preferred wavelength of $l_0 = 2\pi/k_0$. It has a growth time $t_0 = 4\kappa/\nu^2$. Thus, small fluctuations in the initial conditions become of order one on the order of the linear growth time $t_0$. Beyond this time the nonlinear term becomes important. It causes a DC "drift" in the solution since $\overline{\partial_t h} = \frac{\lambda}{2}\overline{(\partial_x h)^2} \neq 0$. Here the overbar denotes spatial average over the entire system with periodic boundary conditions. Without loss of generality, we consider $\lambda > 0$ throughout the text. The time averaged drift rate,

$$v_L = \langle \overline{\partial_t h} \rangle \qquad (2.1)$$

is then a finite positive constant for a system of size $L$, i.e. the solution drifts "upward" to larger values of $h$. The large scale structure, or roughness, of the interface profile originates from fluctuations in the *local* drift rate about its average, $v_L$. Note that the three parameters in Eq. (1.1) define the basic space, time, and height scales, $l_0, t_0$, and $h_0 = 2\nu/\lambda$ respectively. The length of the system $L/l_0$ is the only dimensionless parameter. In the hydrodynamic limit $L/l_0 \to \infty$, the KS equation can be considered *parameter free*.

Consider first small systems with periodic boundary conditions where the drift rate is spatially uniform so that the dynamics are simple. For a system of size $L < l_1 \approx 2\pi/k_1$, there are no unstable modes, so the trivial solution $h = $ Const is a global attractor. For $L \gtrsim l_1$, a nontrivial stationary, cellular solution $H(x; L)$ appears, with $L$ parameterizing the dependence of the solution on the system size. The cellular solution is defined by the ordinary differential equation,

$$w_0(L) = -\nu \frac{d^2 H}{dx^2} - \kappa \frac{d^4 H}{dx^4} + \frac{\lambda}{2}\left(\frac{dH}{dx}\right)^2, \qquad (2.2)$$

in a periodic box of size $L$. The drift rate $w_0(L)$ is explicitly space and time independent for these solutions. It is found [14,16] that for $l_1 < L < l_2 \approx 1.27 l_0$, Eq. (2.2) admits a unique uni-cellular solution which is the global attractor. For $L > l_2$, multi-cellular solutions are possible. The KS equation undergoes a complicated bifurcation sequence with regions of stable multi-cellular solutions between regions of chaos as $L$ is increased beyond $l_2$ [14,15]. The basins of attractions of the stable solutions become vanishingly small for $L \gtrsim 6l_0$.

The stability of the stationary cellular solutions in the region $l_1 < L < l_2$ have been investigated in Ref. [16]. (Ref. [16] actually studied the stability of the velocity field $u = \partial_x h$, but the results carry over directly to the potential field $h$ considered here.) The precise shape of the cells depend on $L/l_0$. Figure 2 shows some examples of the cellular solution. For $L \approx l_0$, one finds that smaller cells have larger amplitudes while larger cells have smaller amplitudes. Since the drift rate

$$v_L = w_0(L) = \frac{\lambda}{2} \int_0^L \frac{dx}{L} \left(\frac{dH}{dx}\right)^2 \qquad (2.3)$$

is a measure of the "non-flatness" of the interface, the dependence of cell shape on $L/l_0$ gives an increasing $w_0(L)$ for decreasing $L$. On the other hand, the cellular solution must vanish at $L = l_1$, so $w_0(L)$ must decrease as $L$ approaches $l_1$. The behavior of $w_0(L)$ is shown in Fig. 3. It has a maximum at $L = l^* \approx 0.85 l_0$. The change in drift rate upon a change of cell size is

$$\sigma(L) = -L \frac{dw_0}{dL}, \qquad (2.4)$$

which changes sign at $L = l^*$. As we will show, $\sigma$ also governs the response of the local drift rate to cell compression in a *large* system, and is crucial in determining the "elasticity" and stability of the large system.

The existence of a continuum of stable, stationary uni-cellular solutions has profound effects on the properties of larger systems. For $L > 2l_1$, stationary cellular solutions comprising of a periodic array of uni-cellular solutions $H(x; l)$ can exist provided that the system size $L$ is commensurate with the cell size $l$. This is not much of a constraint when there exist a broad continuum of $l$'s. Consequently, a degeneracy of periodic cellular solutions can exist for systems of lengths as small as a few $l_0$'s. For example, an $L \approx 4l_0$ system can fit four cells with size $l \approx l_0$ or five cells of size $l \approx 0.8 l_0$. If we evolve the KS equation starting from *random* initial conditions, the system becomes frustrated attempting to choose between the four-cell or five-cell state. Such "frustration" effects can lead to chaotic dynamics [21], since it involves nonlinear coupling of at least three degrees of freedom: the amplitudes of the four-cell state, five-cell state, and their relative phase. The chaotic oscillations between the four-cell and five-cell state then lead to the "creation" and "annihilation" of cells, and are perceived as "space-time defects" (Fig. (IV). Thus the dynamics (or statistical properties) of the defects can be obtained through an investigation of the small chaotic systems. This may be carried out using amplitude equations [21,22] or by direct numerical simulations of small systems. Here we shall assume that the relevant statistical properties of the small chaotic systems are already known, and derive the effective equation of motion describing the KS dynamics for $L \gg l_0$.

## III. PERTURBATIONS ABOUT THE CELLULAR SOLUTIONS

The effective hydrodynamics describing perturbations of the *cellular* solutions are derived first. They have been investigated by Frisch, She, and Thual [16] formally using a multiple-scale perturbation analysis. A similar study was done by Shraiman [7]. Here we shall recover these results from a simple coarse-graining procedure which highlights the underlying physics. Key concepts and methods developed here will be carried over directly into the chaotic regime in Sec. IV. Appreciating the physics of the cellular state will be crucial towards developing a theory for the chaotic state.

We consider the effect of long-wavelength perturbations of the cellular solution $H(x; \ell)$ with period $l_1 < \ell < l_2$ so that the cellular solution is locally stable. The $h$-field is divided into a "slow" part $h_<$ and a "fast" part $h_>$, where slow and fast correspond to long and short spatial distances with respect to $\ell$. More generally, we define

$$[\mathcal{O}]_<(x,t) \equiv \frac{1}{\ell} \int_x^{x+\ell} dy \, \mathcal{O}(y,t), \tag{3.1}$$

$$\text{and} \quad [\mathcal{O}]_>(x,t) \equiv \mathcal{O}(x,t) - \mathcal{O}_<(x,t), \tag{3.2}$$

where the operator $\mathcal{O}$ is a functional of $h$. We shall be interested in the *long wavelength* fluctuations in $h_<$, for which

$$\partial_x^n [h]_< \approx [\partial_x^n h]_< \quad \text{and} \quad [h_>]_< \approx 0. \tag{3.3}$$

Spatially averaging Eq. (1.1) over $\ell$ and applying the approximations stated in Eq. (3.3), we obtain the equation of motion for the slow modes,

$$\partial_t h_< = -\nu\partial_x^2 h_< - \kappa\partial_x^4 h_< + \frac{\lambda}{2}(\partial_x h_<)^2 + w_<(x,t), \tag{3.4}$$

where

$$w_<(x,t) \equiv \frac{\lambda}{2}\int_x^{x+\ell} \frac{dy}{\ell}(\partial_y h_>)^2 \tag{3.5}$$

gives the coarse-grained *local* drift rate.

To complete the coarse-grained dynamics, we need to obtain the equation of motion for $w_<$. This can be derived formally by combining Eq. (3.5) with the equation of motion for the fast variable $h_>$,

$$\partial_t h_> = -\nu\partial_x^2 h_> - \kappa\partial_x^4 h_> + \frac{\lambda}{2}(\partial_x h_>)_>^2 + \lambda\partial_x h_< \partial_x h_>. \tag{3.6}$$

Equation (3.6) is obtained by subtracting Eq. (3.4) from Eq. (1.1) since $h = h_< + h_>$.

However, to emphasize some essential concepts that will be useful later we provide a physically motivated derivation. Consider fast modes comprised of slight local distortions of a cellular solution $H(x;\ell)$, where $\ell$ parameterizes the shape and size of the solution. This is implemented through the use of the ansatz [23]

$$h_>(x,t) = H(x; l(x,t)), \tag{3.7}$$

where $l(x,t)$ is the local cellular wavelength which can deviate slightly from $\ell$. This ansatz with Eqs. (2.3) and (3.5) imply [24]

$$w_<(x,t) = w_0(l(x,t)), \tag{3.8}$$

i.e. $w_<$ only depends on the local fluctuations of cell size $l(x,t)$, or alternatively on the cellular wavenumber, $k(x,t) \equiv 2\pi/l(x,t)$. Hence the time dependence of $w_<$ is given by

$$\frac{\partial w_<}{\partial t} = \sigma(2\pi/k)\frac{1}{k}\frac{\partial k}{\partial t}, \tag{3.9}$$

where $\sigma$ is the response of the single-cell drift rate to a small change in cell size as given in Eq. (2.4).

The dynamics of $k(x,t)$ can be obtained from symmetry considerations. For slight distortions around the cellular solution, the wavenumber must be *locally conserved*. (Nonconservation must involve the creation or annihilation of cells which is excluded in the cellular state by definition.) In addition, the inherent Galilean invariance of the KS equation must be respected. From these considerations, it follows that to leading order the equation of motion of $k(x,t)$ must have the form

$$\partial_t k(x,t) = \lambda \partial_x h_< \partial_x k(x,t) + \lambda k(x,t) \partial_x^2 h_< + \beta \partial_x^2 h_< + \gamma \partial_x^2 k. \tag{3.10}$$

Note that the local conservation law for $k$ excludes a term of the form $-\alpha k(x,t)$ in Eq. (3.10). Comparing the second and third terms on the right hand side of Eq. (3.10), we find the constant $\beta$ to have the dimensions of $\lambda \cdot k$. However, there is no preferred value of $k(x,t)$ in the perturbation theory [25], so we must have $\beta = 0$. The constant $\gamma$ involves the cell height and is generically nonzero; its computation is more complicated and is not discussed here.

To make the slow mode equations more transparent, consider $k(x,t) = (2\pi/\ell)[1 + \delta(x,t)]$, where $\delta(x,t)$ describes a small compression or dilation of the cellular solution of length $\ell$. Clearly, this new variable $\delta(x,t)$ is proportional to the local cell density. In terms of $\delta(x,t)$, the slow mode equations (3.4) and (3.10) becomes

$$\partial_t h_< = -\nu \partial_x^2 h_< + w_0(\ell) + \sigma(\ell)\delta(x,t) + O\left[\partial_x^4 h_<, (\partial_x h_<)^2\right], \tag{3.11}$$

$$\partial_t \delta = \lambda \partial_x^2 h_< + \gamma \partial_x^2 \delta, + O\left[\partial_x \delta \partial_x h_<, \delta \partial_x^2 h_<\right], \tag{3.12}$$

where $\sigma(\ell)$ is the drift rate response. Although the above equations are derived in a phenomenological way, they can be obtained more systematically from a multiple scale analysis, as was done in Ref. [16].

Equations (3.11) and (3.12) give a physical description of the slow mode dynamics. According to Eq. (3.12), cells slide into valleys ($\partial_x^2 h_< > 0$) and thereby become compressed ($\partial_t \delta > 0$), and similarly slide off of hilltops ($\partial_x^2 h_< < 0$) and become dilated ($\partial_t \delta < 0$). This behavior is a straightforward consequence of Galilean invariance as illustrated in Fig. 4.

Fluctuations in the local cell density affect the height profile through a fluctuating local drift rate $\sigma\delta$. For $\sigma > 0$ in Eq. (3.11), cell compression in the valleys leads to a faster local drift rate pushing up the valleys, while cell dilation on the hilltops leads to a slower local drift, pushing down the hilltops.

The above dynamics give rise to oscillations. If the terms with $\nu$ and $\gamma$ (and all higher order ones) are neglected, then Eqs. (3.11) and (3.12) are easily combined to yield the wave equation

$$\partial_t^2 h_< = c^2 \partial_x^2 h_<, \qquad (3.13)$$

where $c^2 = \lambda\sigma(\ell)$. (It can be shown that the expression for $c^2$ given in Ref. [16] can be transformed to this form.) Thus for $\sigma > 0$, the cellular solution is elastic and supports wave propagation. From Fig. 3, we see that this occurs for $\ell \gtrsim l^*$. For smaller $\ell$'s, $c^2 < 0$ and the cellular solution is unstable to small perturbation. Thus for $\sigma < 0$, slower drifting smaller cells eventually become annihilated by their larger neighbors.

In the regime $c^2 > 0$, the oscillation is weakly damped if $\gamma > \nu$, leading to the behavior of visco-elasticity first found in Ref. [16]. However the oscillation becomes unstable if $\gamma < \nu$. This was found to occur for $\ell > 0.91 l_0$ [16]. Thus stable visco-elasticity occurs for a narrow window of cell sizes, $l^* < \ell < 0.92 l_0$.

We re-emphasize that stability of the large scale dynamics requires *elasticity* and *damping*. Elasticity is governed by $\sigma$, which characterizes the response of the local drift rate to a small local compression. The analysis of this section establishes the (somewhat surprising) link between this response and the $\sigma$ describing the size dependence of the single-cell drift rate. This is the case because changing the size of a single, periodic cell is an effective way of mimicking local compression in the cellular state of a large system. It is an example of how large scale dynamics can be predicted by computing some properties of a small, periodic system. We utilize this idea in Sec. IV for the chaotic regime.

While an elastic response exists for a large range of parameters, damping is found to be weak even in the stable regime $\gamma > \nu$. In particular, there is no diffusive relaxation as expected of KPZ-type dynamics. This is a consequence of cell-conservation which excludes a DC-damping term of the form $-\alpha k$ in Eq. (3.10). Note that stability of the cellular state is possible if a damping term $-\mu h$ is included in Eq. (1.1). Although such a term breaks the translational symmetry $h \to h + \text{Const}$ and is not allowed in the KS equation, it does describe a number of other interesting physical phenomena [26] and will be investigated elsewhere. Another possible route to stability is to break the $k$-conservation. This requires cell creation and annihilation occurring in the chaotic state.

## IV. SPATIOTEMPORAL CHAOS

The concepts developed for the perturbed dynamics of the cellular solutions provide a launching point for analyzing STC. As argued in Ref. [7], the system rarely ever settles into a cellular state if it starts from a generic set of initial conditions. Instead, it exhibits chaotic dynamics with inhomogeneous spatial structures as shown in Fig. 1. Chaos occurs for systems of sizes as small as $3l_0 \sim 4l_0$ (see Refs. [14] and [15] for an in-depth discussion of the chaotic dynamics in small systems). The origin of chaotic motion in small systems has already been alluded to in Sec. II: The system is "frustrated" by the existence of multiple "metastable" cellular solutions.

The chaos in a small KS system can be described in terms of its statistical properties given by the probability distribution $\mathcal{P}[H]$, where $H(x,t;\ell)$ is obtained from Eq. (1.1) in a periodic box of size $\ell$ with random initial conditions. In this section, we shall assume that $\mathcal{P}[H]$ is already known for $\ell$ of the order of a few $l_0$'s [14,15], and use it to obtain the statistical properties of the solutions in systems of sizes $L \gg \ell$. In Sec. V, we will numerically measure these statistical properties. It may be possible to attain some analytical understanding of the small system via amplitude equations [21]. The basic premise is that the STC of a large system can be understood in terms of the chaos observed in (mutually coupled) small systems. The size of the small system $\ell$ should be large enough to support a continuum of chaotic states. Let the shortest $\ell$ satisfying this criterion be $l_c$. We find in Sec. V that the procedure worked well for $\ell > l_c \approx 4l_0$.

We wish to understand which properties of the small system are important, and how the small systems are coupled together. We accomplish this with a coarse-graining procedure similar to that used in Sec. III, dividing the $h$-field into "slow" and "fast" parts, i.e. $h = h_< + h_>$, with respect to $\ell$. The equation of motion for $h_<$ and $h_>$ are again Eqs. (3.4) and (3.6). It is important to note that the dynamics of the slow modes affect those of the fast modes through the last term in Eq. (3.6).

To complete the description of slow mode dynamics, we need to specify the probability distribution of $w_<$, obtained from the distribution of $h_>$ through the definition Eq. (3.5). As in the cellular case, we would like to relate $h_>$ to $H$, the solution in a small, periodic system. However, generalization of the simple ansatz (3.7),

$$h_>(x,t) = H(x,t;l(x,t)), \tag{4.1}$$

no longer applies here, because a) "wavelength" is no longer well defined in the chaotic case, and b) the chaotic dynamics depend only weakly on the system size (provided the size is large enough to support the chaotic state). In particular, we can no longer mimic a local compression in a large system by changing the size of a small periodic system as was done in Sec. III.

In the chaotic case, the simplest ansatz possible is

$$\text{Prob}[h_>(x,t)] = \mathcal{P}[\mathcal{H}(x,t;\ell); \partial_x^2 h_<] \tag{4.2}$$

where $\mathcal{H}$ is the solution of Eq. (3.6) in a small, periodic system of size $\ell$ under the influence of a slow imposed curvature $\partial_x^2 h_<$. The explicit dependence of $\mathcal{P}$ on the slow mode $h_<$ comes from the last term of Eq. (3.6). (Since a constant $\partial_x h_<$ merely amounts to a Galilean boost, the leading nontrivial behavior comes from the dependence on $\partial_x^2 h_<$.) Note that in using periodic boundary conditions for $\mathcal{H}$, this ansatz neglects spatial couplings of $h_>$ between the neighboring coarse-grained regions. In this and the following sections, we will first derive the consequence of the ansatz (4.2), and then show self-consistently that the spatial couplings neglected do not alter the hydrodynamic behavior.

Given the ansatz (4.2), the probability distribution of $w_<$ becomes

$$P[w_<; \partial_x^2 h_<] = \int \mathcal{D}\mathcal{H}\, \delta\left(w_< - \frac{\lambda}{2}\overline{(\partial_x \mathcal{H})^2}\right) \mathcal{P}[\mathcal{H}; \partial_x^2 h_<], \qquad (4.3)$$

i.e. it becomes an explicit functional of the chaotic solution of a small KS system, and can therefore be obtained numerically (see Sec. V) or possibly even analytically. Moments of the distribution $P$ can then be used to construct an appropriate stochastic equation for $w_<$. A review of this construction is given in the Appendix. Here, we shall motivate the form of the effective equation of motion for $w_<$.

From the definition of $w_<$ in Eq. (3.5), we have

$$\partial_t w_<(x,t) = \lambda [\partial_x h_> \partial_x (\partial_t h_>)]_<, \qquad (4.4)$$

where $[\cdot]_<$ denotes spatial average as in Eq. (3.2). Substituting Eq. (3.6) for the dynamics of the fast modes leads to

$$\partial_t w_< = F[w_<] + \lambda^2 [\partial_x h_> \partial_x (\partial_x h_< \partial_x h_>)]_<, \qquad (4.5)$$

where all the purely fast mode contributions are represented by $F[w_<]$. To find the form of $F[w_<]$, consider first the simple case where the system size is small, $L = \ell \gtrsim l_c$, so that the second term in Eq. (4.5) vanishes (i.e. no slow mode contributions). In this case, $w_<(t) \equiv \overline{w_<(x,t)}$ is just the instantaneous, spatially-averaged drift rate of the small chaotic system, fluctuating about its time-averaged value, $\langle w_<(t) \rangle = v_\ell$. It is then natural to expect the dynamics of $w_<$ to be described by the following Langevin equation,

$$\partial_t w_<(t) = -\alpha_\ell (w_<(t) - v_\ell) + \xi(t), \qquad (4.6)$$

to leading order in $w_< - v_\ell$, with a response time $\alpha_\ell^{-1}$ and a Langevin noise $\xi(t)$. We will show in the Appendix that the numerical result of Sec. V leads to Eq. (4.6) as a correct description of the leading moments of $P[w_<]$. The parameter $\alpha_\ell$, $v_\ell$ and the second moment of the noise

$$\langle \xi(t)\xi(t') \rangle = \Delta_\ell(t - t') \qquad (4.7)$$

are determined numerically in Sec. V.

The second term in Eq. (4.5) vanishes as well in a large system with $L \gg l_c$ if we set $h_< = $ Const. In this case, we may expect the dynamics of $w_<$ to be similar to Eq. (4.6), but with a space-dependent $w_<$ and $\xi$, i.e.

$$\partial_t w_<(x,t) = F[w_<] = -\alpha_\ell(w_<(x,t) - v_\ell) + \xi(x,t). \tag{4.8}$$

Note that spatial coupling terms such as $\partial_x^2 w_<$ should generically be present in Eq. (4.8). These are precisely the couplings neglected by the use of ansatz (4.2). They will be shown to be *irrelevant* in the hydrodynamic limit and will not be pursued here. In the simplest scenario, we expect the Langevin noise $\xi(x,t)$ to be uncorrelated in space for spatial scales $\gg l_c$, i.e.

$$\langle \xi(x,t)\xi(x',t') \rangle = \delta(x-x')\tilde{\Delta}_\ell(t-t'). \tag{4.9}$$

If this is true then we would expect the spatial average of $\xi(x,t)$ (i.e. $\overline{\xi}_L(t)$), to be given by the following correlator,

$$\langle \overline{\xi}_L(\tau)\overline{\xi}_L(0) \rangle = \Delta_L(\tau) = \tilde{\Delta}_\ell(\tau)/L, \tag{4.10}$$

for $L$ exceeding several $l_c$'s. This $1/L$-dependence will be verified by the numerics in Sec. V.

Next we consider the effect of the second term on the right hand side of Eq. (4.5). It can be rewritten as

$$\lambda^2[(\partial_x h_>)^2 \partial_x^2 h_<]_< + \frac{\lambda^2}{2}[(\partial_x h_<)\partial_x(\partial_x h_>)^2]_<$$
$$= 2\lambda w_< \partial_x^2 h_< + \lambda \partial_x h_< \partial_x w_< + \lambda [x \partial_x w]_< \partial_x^2 h_<. \tag{4.11}$$

The dominant term is obtained by approximating $w_<$ by its time average $v_\ell$. Then to lowest order Eq. (4.11) becomes $2\lambda v_\ell \partial_x^2 h_<$. Higher order terms are again irrelevant as will be shown.

Putting the above together, and re-expressing in terms of the local drift rate fluctuation,

$$\pi(x,t) = w_<(x,t) - v_\ell, \tag{4.12}$$

we obtain the equations

$$\partial_t h_< = -\nu \partial_x^2 h_< + \frac{\lambda}{2}(\partial_x h_<)^2 + v_\ell + \pi(x,t) + O\left[\partial_x^4 h_<\right] \tag{4.13}$$

$$\partial_t \pi = -\alpha_\ell \pi + \beta_\ell \partial_x^2 h_< + \xi(x,t) + O\left[\partial_x^2 \pi, \partial_x \pi \partial_x h_<, \pi \partial_x^2 h_<, \pi^2, (\partial_x^2 h_<)^2\right], \tag{4.14}$$

describing the coarse-grained dynamics, with $\beta_\ell = 2\lambda v_\ell$. A formal construction of Eq. (4.14) based on the numerical results of Sec. V is presented in the Appendix.

It is useful to compare Eqs. (4.13) and (4.14) to the cellular slow mode equations (3.11) and (3.12). We note that the drift rate fluctuation $\pi(x,t)$ resembles the cell density fluctuation $\delta(x,t)$. This should be expected: For any quasi-cellular profile $h$, the local drift rate operator $w_<(x,t) = \left[\frac{\lambda}{2}(\partial_x h_>)^2\right]_<$ has two peaks in $x$ for each cell, therefore the long wavelength part of $w_<(x,t)$ should provide a measure of the local cell density (at least in one spatial dimension). Identifying $\pi(x,t)$ with the cell density fluctuation, we see that as in the cellular case, the chaotic dynamics contains elasticity. Cells are compressed ($\partial_t \pi > 0$) in valleys ($\partial_x^2 h_< > 0$) and dilated ($\partial_t \pi < 0$) on hilltops ($\partial_x^2 h_< < 0$). Compressed (dilated) regions have higher (lower) drift rates, thus providing a restoring force to height fluctuations. The elasticity is again quantified by the drift response of the system to compression. In the cellular case, compression was effectively mimicked by changing the cell size, but in the chaotic case, compression is produced by explicitly imposing an external curvature $\partial_x^2 h_<$ as described in Sec. V. (Defining $\sigma \delta = \pi$, we see that the drift response $\sigma = 2v_\ell$ is always positive in the chaotic case.) The main additions due to the chaotic dynamics are the damping term $-\alpha_\ell \pi$ and the noise term $\xi(x,t)$ in Eq. (4.14). As alluded to earlier, the damping term is allowed in the chaotic dynamics because cell density is no longer a conserved quantity (there is a preferred $k$), while the noise $\xi(x,t)$ describes the stochastic nature of the cell creation and annihilation process. In fact, Eq. (4.14) is the minimal equation demanded by symmetry in the chaotic regime.

The damping term makes the chaotic dynamics fundamentally different from the cellular dynamics. In the hydrodynamic limit, Eq. (4.14) can be solved to yield

$$\pi(x,t) = \frac{\beta_\ell}{\alpha_\ell}\partial_x^2 h_< + \frac{\xi}{\alpha_\ell} + O\left[\partial_x^4 h_<, \partial_x^2\partial_t h_<, (\partial_x^2 h_<)^2\right]. \tag{4.15}$$

Substituting Eq. (4.15) into the slow mode equation (4.13), we finally obtain the effective equation

$$\partial_t h_< = \tilde{\nu}_\ell \partial_x^2 h_< + \frac{\lambda}{2}(\partial_x h_<)^2 + v_\ell + \eta(x,t) + O\left[\partial_x^4 h_<, \partial_x^2\partial_t h_<, (\partial_x^2 h_<)^2\right], \tag{4.16}$$

with

$$\tilde{\nu}_\ell = \frac{\beta_\ell}{\alpha_\ell} - \nu, \tag{4.17}$$

$$\langle \eta(x,t)\eta(x',t')\rangle = \delta(x-x')\widetilde{\Delta}_\ell(t-t')/\alpha_\ell^2. \tag{4.18}$$

As will be shown in Sec. V, $\widetilde{\Delta}_\ell(\tau)$ is correlated only up to a time scale $\tau_0 \sim t_0$. Thus for time scales $\gg \tau_0$ (in the hydrodynamic limit), it may be approximated by a $\delta$-function. Hence

$$\langle \eta(x,t)\eta(x',t')\rangle = \widetilde{D}_\ell \delta(x-x')\delta(t-t'), \tag{4.19}$$

with an effective noise amplitude

$$\widetilde{D}_\ell = \int_{-\infty}^{\infty} d\tau\, \widetilde{\Delta}_\ell(\tau)/\alpha_\ell^2. \tag{4.20}$$

The "bare" parameters $\tilde{\nu}_\ell, \widetilde{D}_\ell$, are determined in Sec. V through simulations of small KS systems of size $\ell$. We find the result $\tilde{\nu}_\ell > 0$ which establishes the hydrodynamic system Eqs. (4.16) and (4.19) to be in the KPZ universality class, provided that the coefficients of the irrelevant terms (those in the square bracket of Eq. (4.16)) are not pathological. In order to compare the results to the parameters $\nu_{\text{eff}}, D_{\text{eff}}$ of the effective KPZ equation measured at very large scales [12], additional renormalization of $\tilde{\nu}_\ell, \widetilde{D}_\ell$ by the irrelevant terms must be considered. This issue will be addressed in Sec. VI.

## V. NUMERICAL METHODS AND RESULTS

The effective hydrodynamic equation given by Eq. (4.16) and Eq. (4.19) must be checked for self consistency by computing the parameters $\tilde{\nu}_\ell$ and $\widetilde{D}_\ell$. One must demonstrate that the effective surface tension $\tilde{\nu}_\ell$ is positive and that the effective noise is spatially and temporally uncorrelated. This is accomplished by making numerical measurements of the leading moments of the probability distribution function $P[w_<; \partial_x^2 h_<]$. As shown in the Appendix and in Sec. IV the moments are directly related to the parameters of the effective equation. The task is greatly simplified by the use of ansatz (4.2) which replaces a small segment of a large system by a small periodic system of the same size subject to some relevant external forcing (curvature). The spatial couplings neglected are recognized as irrelevant once Eq. (4.16) is established to described the hydrodynamic limit.

In order to compare consistently to the results of the large KS system obtained in Ref. [12], we used the same numerical scheme: A simple real space finite differencing scheme was used with the nonlinearity discretized as $\frac{\lambda}{2}[(h_{i+1}(t) - h_{i-1}(t))/2]^2$, where $i$ is the discretized spatial index. An Euler step was used for the time evolution, with step size $dt = 0.1$. The simulations were done with the same parameters as those in Ref. [12], namely $\nu = 1, \kappa = 1, \lambda = 2$, for which $l_0 \simeq 8.89$ and $t_0 = 4$. The measurements were made on systems ranging in sizes from $L = 32$ to $L = 150$, corresponding to $3.6 l_0$ to $16.9 l_0$. Simulations were performed on an IBM RS6000 workstation unless otherwise specified. It should be noted that the lattice KS equation studied in Ref. [12] and here is actually a dynamical system *distinct* from the continuum KS equation (1.1). Although we find that the lattice KS equation belongs to the same universality class as the continuum equation, eventually flowing to the same KPZ fixed point, the precise values of the "bare" parameters of the resulting effective hydrodynamic equation (4.16) do depend on numerical and discretization schemes. However, since our approach is not very computationally demanding, a careful, high accuracy numerical method could be easily employed to get values that better represent the continuum KS equation.

As shown in the Appendix, it is necessary to obtain the first and second moments of the distribution $P[w_<; \partial_x^2 h_<]$, as well as the response function. In addition, due to the discretization scheme used, Galilean invariance is broken at the microscopic scale. This leads to a modified response to uniform tilt $\partial_x h_<$ (a renormalized nonlinear parameter $\lambda_{\text{eff}}$) which is also investigated.

*1. Response Function and Stability*

The effective surface tension may be obtained from the time-dependent response of the drift rate $w_<$ to an imposed curvature $\partial_x^2 h_<$ in small systems of the size of several $l_0$'s (see Appendix). Recall from Galilean invariance that a slow curvature compresses or dilates the cells (see Fig. 4), thereby changing the drift rate. It is this change of drift rate in the chaotic state that is desired. The measurement was implemented numerically by simulating the dynamics of the fast variable Eq. (3.6) with a slow curvature imposed through the choice of $h_<$.

Ideally a uniform forcing $\partial_x^2 h_< = \text{Const}$ (i.e. $h_< \propto x^2$) should be applied. However, in practice an infinitely large parabola can never be used and boundary effects will become important. The drift response will not be uniform at the boundaries thus distorting measurements. This problem may be overcome in a number of ways. One could use a large enough parabola and only look at the drift response near the center away from the boundaries. However, depending on the length scale over which the response is probed, this could require a rather large system and become computationally intensive. Another way is to apply a slowly varying $\partial_x^2 h_<$ (e.g. a sine wave), systematically remove the nonuniform effects and measure the response to the average curvature imposed. We choose the second method.

With a spatially nonuniform forcing Eqs. (4.2) and (4.3) must be slightly modified. One should use $w_< = \frac{\lambda}{2}\overline{(\partial_x \mathcal{H}_>)^2}$, where $\mathcal{H}_>$ comprise of the Fourier modes that are "fast" compared to the spatial variation in $\partial_x^2 h_<$. The corresponding "slow" modes $\mathcal{H}_<$ are a "back-reaction" to the nonuniform curvature and must be subtracted away. The back-reaction arises because each region experiencing a different curvature will have a different response. This will then distort the surface and alter the effect of the imposed curvature. To ensure that the back reaction has been properly removed the time averaged spatial profile must be monitored. The criterion for effective back reaction neutralization is that the time average of the spatial profile remains flat. i.e. $\langle \partial_x \mathcal{H}(x,t) \rangle = 0$.

A sine wave was specifically chosen for the slow forcing because the shape of the "back reaction" will also be a sine wave to linear order and can be easily removed. However, there are still nonlinear effects that can alter the background curvature. If significant spatial nonuniformity in $\langle \mathcal{H} \rangle$ still exists after the first Fourier mode is removed, then higher modes are also removed until the surface is uniform to an acceptable tolerance ($< 10\%$ of the forcing amplitude). For the system sizes used we never needed to remove more than two modes.

In the simulation, Eq. (3.6) was used with

$$h_<(x) = -c' \left(\frac{L}{\pi}\right)^2 \sin\left(\frac{\pi x}{L}\right), \qquad \text{for} \qquad 0 \leq x < 2L. \tag{5.1}$$

Hence the left and right halves of the system were under the influence of a positive and a negative curvature respectively. In each run, two measurements of the drift rate, one for the left and one for the right half were made. We took the imposed curvature $c$ to be the 'average' curvature experienced by each half,

$$c = \frac{1}{L} \int_0^L \partial_x^2 h_< \, dx = \frac{c'}{\pi}. \tag{5.2}$$

The first Fourier mode of $h_>$, $h_>(k = \pi/L)$, was removed after each time step. This was accomplished by using an FFT, zeroing the first mode, then using an inverse FFT to restore to real space. The DC mode was also zeroed to remove the average drift component. The second mode was removed if necessary.

The measurement was performed by starting the simulation with $c' = 0$ and with small random initial conditions. After the system settled into the chaotic attractor (several linear growth times), a small curvature ($c' > 0$) was imposed at $t = 0$ and the new drift rate was measured by taking $w_<(t) = \frac{\lambda}{2}[\frac{1}{L}\sum(\partial_x h_>)^2]$ separately for the left and right halves. The measurement was then averaged over many configurations of random initial conditions for each $h_<$ used. Our main measurement was for $L = 64$ averaged over $10^5$ samples and for $c'$ ranging from 0 to 0.004. This corresponded to $L \sim 7l_0$ which was large enough to capture the essential small system chaotic behavior yet small enough to be computed comfortably.

The spatial nonuniformity after the removal of the first Fourier modes was found to be under 10% of the amplitude of $h_<$. We also made measurements for $L = 32$ and $L = 128$ but with fewer samples and a smaller range of $c'$.

Fig. 5 shows a typical form of $\langle w_<(t) \rangle$ for $c' = 0.002$ and $L = 64$ for the two halves of the system. The response rises/drops sharply before eventually saturating to a constant value. This behavior is well described by the exponential form

$$\langle w_<(t) \rangle = v_L + A_L(c') \left( 1 - e^{-\alpha_L(c')t} \right). \tag{5.3}$$

The values of $A_L(c')$ and $\alpha_L(c')$ were computed. We find

$$\alpha_L = 0.15 \pm 0.02 \tag{5.4}$$

to be independent of $c'$, while $A(c')$ increases monotonically as shown in Fig. 6. A least-square fit finds

$$A_L(c') = (4.6 \pm 0.2)c' + (200 \pm 15)(c')^2. \tag{5.5}$$

The effect of the nonlinear term is small over the range of $c'$ used. From the exponential form of the linear response Eq. (5.3) we obtain (see Appendix) the effective equation of motion for $w_<$,

$$\partial_t w_< = -\alpha_L(w_< - v_L) + \beta_L c + \xi(x, t), \tag{5.6}$$

where $c \equiv \partial_x^2 h_<$ is the average slow curvature forcing. We can identify

$$c \frac{\beta_L}{\alpha_L} = A_L(c'), \tag{5.7}$$

which leads to

$$\frac{\beta_L}{\alpha_L} = \pi \cdot \frac{d}{dc'} A_L(c') \bigg|_{c'=0} = 14.5 \pm 0.6, \tag{5.8}$$

using $c' = \pi c$ from Eq. (5.2). This gives an effective surface tension of

$$\tilde{\nu}_L = \frac{\beta_L}{\alpha_L} - \nu = 13.5 \pm 0.6. \tag{5.9}$$

These measurements have been repeated for different $L$'s. For $L = 32$ we found the values: $\alpha_L = 0.15 \pm 0.03$, $\tilde{\nu}_L = 12.5 \pm 1$. For $L = 128$, removing the first Fourier mode was not enough to counter the back reaction. We removed the first two modes and obtained the results: $\alpha = 0.14 \pm 0.03$ and $\tilde{\nu}_L = 14 \pm 2$. The larger uncertainties are due to the fact that fewer samples and a smaller range of $c'$ was used in the measurements. We find only a weak $L$-dependence for the parameters $\alpha_L$ and $\beta_L$ as long as the small KS system is well within the chaotic state and the back reaction is removed.

We have also tried a different forcing term consisted of two parabolas of opposite curvatures:

$$h_<(x) = \begin{cases} \frac{c}{2}\left(x - \frac{L}{4}\right)^2 - \left(\frac{L}{2}\right)^2 & \text{for} \quad 0 < x < \frac{L}{2} \\ -\frac{c}{2}\left(x - \frac{3L}{4}\right)^2 + \left(\frac{L}{2}\right)^2 & \text{for} \quad \frac{L}{2} < x < L \end{cases} \quad (5.10)$$

instead of the sine forcing Eq. (5.1). In this case the back reaction was removed by removing the first mode and/or removing the average drift of the right and left halves. We find results similar to those reported above, except with larger uncertainties. This is attributed to the fact that our procedure of removing the back reaction was incomplete even to linear order.

### 2. Noise Amplitude

The effective noise amplitude was calculated from the two-point correlation function of the drift rate $w_<(t)$,

$$C_L(\tau) = \langle (w_<(t + \tau) - w_<(t))^2 \rangle, \quad (5.11)$$

for a small system with no external forcing and with periodic boundary conditions. The numerical scheme used was the same as that for the response function. Starting from random initial conditions and after allowing the system to reach the steady state, $C_L(\tau)$ was measured up to a time when the correlation function saturates. The result was then averaged over $10^4$ configurations of random initial conditions. A typical form of the averaged correlation function is shown in Fig. 7 for $L = 100$. From Eq. (5.6) with $c = 0$ and for intrinsic noise $\xi(x,t)$ that is completely uncorrelated in space and time, $\langle \xi(x,t)\xi(x',t')\rangle = B\delta(x-x')\delta(t-t')$, then

$$C_L(\tau) = \frac{B}{L\alpha_L}\left(1 - e^{-\alpha_L \tau}\right). \tag{5.12}$$

The fine structure in the numerically obtained $C_L(\tau)$ shown in Fig. 7 indicates short-time structure in $\xi(t)$, which can be obtained by de-convolving $C_L(\tau)$ from the linear response function $G(\tau)$ as in Eq. (A18). However, this short-time structure does not affect the hydrodynamic limit, and is not pursued. As explained in Sec. IV, the quantity relevant to hydrodynamic behavior is the time integral of the correlator

$$D_L(\tau) = \langle (w_<(t) - v_L)(w_<(t+\tau) - v_L)\rangle = \frac{1}{2}\left(C_L(\infty) - C_L(\tau)\right). \tag{5.13}$$

If $\xi$ is short-range correlated in space, we expect

$$\widetilde{D}_L = L\int_{-\infty}^{\infty} d\tau\, D_L(\tau) \tag{5.14}$$

to be independent of $L$. This quantity has been measured for various system sizes ranging from 50 to 150. We find

$$\widetilde{D}_L = 2.6 \pm 0.1, \tag{5.15}$$

with negligible $L$-dependence (see Fig. 8).

### 3. Breaking of Galilean Symmetry

In Ref. [12] it was found that the breaking of Galilean symmetry by finite differencing resulted in an increase of $\lambda$ from the bare value of 2 to $4.69 \pm 0.15$ for a large KS system with $L = 1024$. We performed the same measurement by determining the dependence of the average drift rate $v_L$ to a macroscopic tilt $\theta$, which is imposed on the system via helical boundary conditions, $h(L) = h(0) + \theta L$. Fitting the data to the parabola $v_L(\theta) = v_L + \frac{\widetilde{\lambda}_L}{2}\theta^2$ we found $\widetilde{\lambda}_L = 4.4 \pm 0.2$ and $v_L = 0.493 \pm 0.001$ for $L = 100$. The drift velocity $v_L$ has a very weak dependence on $L$. As in Ref. [12], we found $\widetilde{\lambda}$ to be sensitive to both the time and spatial discretization size, with $\widetilde{\lambda}_L$ approaching $\lambda$ as the discretization step sizes are made smaller.

Another way of determining the effect of Galilean symmetry breaking is through a measurement of the *advection rate* in the presence of a macroscopic tilt. For the continuum system, Galilean invariance gives $x \to x + ut$, with the advection rate $u = \lambda\theta$ if $h \to h + \theta x$. However, since our discrete system breaks Galilean invariance, it is not *a priori* obvious what $u(\theta)$ should be in this case. In the chaotic state, one can measure $u(\theta)$ through the average response function $\langle \mathcal{G}(x,t) \rangle$, which can be obtained by following the difference in time evolution of two samples with identical (generic) initial conditions except for a small difference at the origin ($x = 0$), and then averaging over all pairs of random initial conditions. For $\theta = 0$, $\langle \mathcal{G}(x,t) \rangle$ is an oscillating function in $x$, peaked at $x = 0$, with decaying amplitude for large $x$ (see Fig. 9). With $\theta \neq 0$, the advection rate can be measured through the movement of the peak of $\langle \mathcal{G}(x,t) \rangle$ in $x$. In practice, however, the above is a very time-consuming task, as a large ensemble average is needed to obtain $\langle \mathcal{G}(x,t) \rangle$ for each $\theta$.

To get an idea of the behavior of the advection rate, we measured the tilt-dependence of our discretized system for the stationary cellular solution. First we checked the tilt-dependence of the drift rate as defined in the previous paragraph. We found $\widetilde{\lambda} = 2.8 \pm 0.2$ for $L = 8$ and $\widetilde{\lambda} = 4.8 \pm 0.2$ for $L = 10$. This shows a strong lattice effect in the cellular solutions. (The values of $\widetilde{\lambda}$ obtained in the stationary and chaotic states are not expected to coincide, since as already described in Sec. III, the drift rate of the stationary state has a strong $L$-dependence even for $\theta = 0$.) Nevertheless, when we measured the advection rate we recovered the result $u = \lambda\theta$ with the *bare* value of $\lambda$ independent of $L$. We believe this result concerning the non-renormalization of the advection rate to hold also in the chaotic state, since the *local* advection rate there should be similar to the size-independent advection rate of the stationary state.

## VI. ANALYSIS OF RESULTS

The existence of a linear response of the drift rate $w_<$ to external curvature $\partial_x^2 h_<$ (Fig. 6) validates Eq. (5.6) as the equation of motion for $w_<$, or equivalently Eq. (4.14) for the drift rate fluctuation $\pi(x,t) = w_<(x,t) - v_L$. This leads to Eq. (4.16) as the effective equation of motion for the slow modes $h_<(x,t)$, with the effective surface tension

$$\widetilde{\nu}_\ell = \frac{\beta_\ell}{\alpha_\ell} - \nu \approx 13.5 \pm 0.6. \tag{6.1}$$

which depends weakly on $\ell$. The positivity of $\widetilde{\nu}_\ell$ along with the independence of $\widetilde{D}_\ell$ on $\ell$ (Fig. 8) establishes Yakhot's conjecture that the KS dynamics belongs to the universality class of the KPZ equation. It then follows that the terms neglected in our treatment, in particular, the spatial coupling terms ($\partial_x^2 \pi$, etc. in Eq. (4.14)) neglected due to the use of the ansatz (4.2), are irrelevant at the KPZ fixed point. We of course cannot *a priori* exclude the possibility that some of the coefficients of the irrelevant terms are very large, e.g. the neighboring chaotic attractors are strongly coupled spatially, thus driving the system to a different fixed point. This possibility can be checked by applying a space-dependent curvature perturbation in Eq. (3.6) and measuring the space-dependent response of the drift rate. The lack of strong $\ell$-dependence in the parameters $\widetilde{\nu}_\ell$ and $\widetilde{D}_\ell$ measured suggests that such effects are not important.

The remainder of this section is devoted to a more quantitative analysis of the numerical results obtained. From the measurement of the saturated drift rate response to external curvature (Eq. (5.5)), and the response time $\alpha_\ell$ (Eq. (5.4)), we obtained $\beta_\ell = 2.2 \pm 0.2$. The analysis of Sec. IV predicts

$$\beta_\ell = 2\lambda v_\ell. \tag{6.2}$$

The drift rate for for $\ell = 64$ is $v_\ell \approx 0.49$. Thus Eq. (6.2) is satisfied within numerical uncertainties if the bare value $\lambda = 2$ is used. There is, of course, an ambiguity as to which is the "proper" $\lambda$ to use for the lattice KS equation studied here, since the effective $\lambda$ obtained from drift and advection response to tilt are different as shown in Sec. V. We argue that the proper $\lambda$ to use is the advection response, $\lambda = 2$. As already explained, the drift rate fluctuation $\pi$ is proportional to cell density fluctuation which responds to a nonzero background curvature through advection. The cell density increases at the bottom of a valley because the advection process brings cells in from the two sides of the valley (see Fig. 4). Thus, the response of cell density to changes in background curvature is determined by the advection rate, which is proportional to the bare $\lambda$. (A similar ambiguity also exists for perturbations about the cellular solutions, where the analysis based on the continuum equation predicts wave dynamics Eq. (3.13) with $c^2 = \lambda \sigma$. For the lattice KS equation studied here [and possibly the pseudo-spectral system studied in Ref. [16] although $\widetilde{\lambda}$ was never measured], it is again unclear as to which is the proper $\lambda$ to be used. However, our numerics [and those of Ref. [16]] clearly support the use of the bare $\lambda$ and are incompatible with the renormalized value $\widetilde{\lambda}$ obtained from the drift response.)

As mentioned in Sec. III, the large scale KS equation is parameter free in terms of the "natural" units, the typical cell size $l_0 = 2\pi\sqrt{2\kappa/\nu} \approx 8.89$, the linear growth time (which is of the order of the maximum Lyapunov time) $t_0 = 4\kappa/\nu^2 = 4$, and the cell height $h_0 = 2\nu/\lambda = 1$. In these units, the bare parameters are $\widetilde{\nu} \approx 0.7\,[l_0^2/t_0]$, $\widetilde{D} \approx 1.2\,[h_0^2 l_0/t_0]$, i.e. they are of order unity. This should be expected from the defect-mediated stability mechanism — both the stability factor $\widetilde{\nu}$ and noise $\widetilde{D}$ are due to the microscopic cell creation/annihilation process which involves the change of the local interface height by an amount $h_0$ over a spatial extent of $l_0$ and is completed within a time $t_0$.

For the same reason, we expect the bare coefficients of the irrelevant terms, (terms in the square brackets in Eq. (4.16)), also to be of the order unity in their respective natural units, at the smallest possible scale $l_c$ at which Eq. (4.16) can be defined. These irrelevant terms will certainly cause finite renormalization of the effective parameters $\widetilde{\nu}_\ell$ and $\widetilde{D}_\ell$ (by an amount of the order unity in natural units), eventually leading to the effective KPZ parameters $\nu_{\text{eff}}$ and $D_{\text{eff}}$ measured in Ref. [12] for coarse-graining scales $\ell \gg l_c$, Thus the parameters $\widetilde{\nu}_\ell = 13.5 \pm 0.6$ and $\widetilde{D}_\ell = 2.6 \pm 0.1$ measured in Sec. V are not quite the macroscopic parameters $\nu_{\text{eff}}$ and $D_{\text{eff}}$ since $\ell = 64$ used is only about twice $l_c$. They serve as an order-of-magnitude estimate of $\nu_{\text{eff}}$ and $D_{\text{eff}}$. Better estimates of the effective parameters can possibly be made by repeating the numerics of Sec. V for larger systems. However, since renormalization by the irrelevant terms quickly saturates for $\ell > l_c$, the weak $\ell$-dependence of the parameters $\tilde{\nu}_\ell$ and $\widetilde{D}_\ell$ found suggests that the range of $\ell$ used is probably already close to the saturated region. Thus the numerical values of $\tilde{\nu}_\ell$ and $\widetilde{D}_\ell$ obtained should be close to $\nu_{\text{eff}}$ and $D_{\text{eff}}$. This is indeed the case, since Ref. [12] finds $\nu_{\text{eff}} = 10.5 \pm 0.6$ and $D_{\text{eff}} = 3.2 \pm 0.1$.

It should be noted that an accurate determination of the bare parameters of the KPZ equation is possible in this case only because the cross-over length scale $l_\times$ beyond which the KPZ nonlinearity becomes important is unusually large. As estimated in Ref. [12],

$$l_\times \approx 150 l_0/\hat{g}, \qquad (6.3)$$

where the dimensionless coupling constant is

$$\hat{g} = \frac{\lambda_{\text{eff}}^2 D_{\text{eff}}}{\nu_{\text{eff}}^3} l_0 \approx 0.53. \qquad (6.4)$$

Since $l_\times \gg l_0 \sim l_c$, it is possible to perform the numerics at $\ell \gg l_c$, and include the effect of the irrelevant operators without running into the complication of renormalization by the *relevant* KPZ nonlinearity. Note that the large value of $l_\times$ is a result of the large prefactor, 150 in Eq. (6.3). This is purely a "KPZ effect". What reflects the inherent nature of the KS dynamics is the dimensionless constant $\hat{g}$ which is of order unity. Of course, a large separation of scales between $l_\times$ and $l_0$ should not be expected to hold for generic STC problems. Without a clear separation of scales, it will not be possible to determine the "bare" parameters of the effective hydrodynamic theory by this or any other methods. (Indeed, the bare parameters will not even be observable in such cases.) Nevertheless, one can still use the approach given here to construct the form of the effective equation, from which the asymptotic, universal behavior may be obtained.

# VII. SUMMARY AND FUTURE OUTLOOK

We have presented an effective hydrodynamic theory of spatiotemporal chaos exhibited by the Kuramoto-Sivashinsky equation. We constructed an effective stochastic equation belonging to the KPZ universality class in the hydrodynamic limit by incorporating the chaotic dynamics of the small KS system in a coarse-graining procedure. This result was established by computing the bare parameters of the effective equation. They are found to be of the same order as the ones obtained from large scale simulations. An important assumption in our approach is weak spatial coupling of the local chaotic dynamics. This allows the use periodic boundary conditions when measuring properties of the small KS systems. The assumption is justified because of the irrelevancy of these spatial couplings at the KPZ fixed point.

Due to Galilean invariance, fluctuations in the background curvature are coupled to fluctuations in the cell density. Thus a preference for zero curvature (positive surface tension) is unavoidably connected to a preference for a specific cell density. The latter is only possible in the presence of spacetime defects, which are manifestations of the division of thick cells and the coalescence of thin cells. We refer to this process as "defect-mediated stability". The dynamics of the defects, which involve the local change of wave number, are chaotic in the KS system because a band of states with nearby wave numbers are metastable and thereby frustrated as discussed in Sec. II. Thus stability is accompanied by stochasticity. In the absence of defects (the cellular case) wavenumber conservation only leads to visco-elastic motion at best.

It is instructive to examine KS-like equations, i.e. equations similar to Eq. (1.1) but with linear operators of the form $|k|^a - |k|^b$, $b > a$. From symmetry, one might again conjecture that these systems belong to the KPZ universality class. However, in light of the analysis presented, the stability mechanism can only produce a term proportional to $-k^2$ in the slow mode equation. Therefore, one can conclude right away that a necessary condition for stability is $a \geq 2$. This explains why instabilities from the linear operator $|k| - k^2$ cannot be tamed by $(\partial_x h)^2$-type nonlinearity, a result well known from the study of the viscous-fingering instability [27]. On the other hand, KPZ behavior is expected for $a > 2$. The bare coefficients can be computed using the scheme of Sec. V and can be compared to those from large scale simulations [28].

If the system is such that the band of stable, single-cell solutions (the analog of the range of sizes between $l_1$ and $l_2$ in Sec. II) is very narrow compared to the typical cell sizes, then the size of the "small" system required to extract local chaotic dynamics must be very large. This is because frustrations due to the coexistence of multi-cellular solutions of different wavenumbers is needed for chaos (recall the four-cell, five-cell example in Sec. II), while the minimum system size $l_c$ required to accomodate different wavenumber states is inversely proportional to the stable band width. It is tempting to conjecture that the density of positive Lyapunov exponents will scale as $1/l_c$ at length scales $\gg l_c$. Such behavior may be investigated in systems exhibiting an order-chaos transition.

It should also be interesting to extend the present approach to study the behavior of the 2+1 dimensional generalization of the KS equation. Again from symmetry, one can make a case for the KPZ equation to describe the hydrodynamic behavior. This approach is advocated by Jayaprakash et al. [29]. On the other hand, fundamentally different behavior is proposed by L'vov and Procaccia [30]. The differences between the two proposals are not expected to be resolved numerically with current computational capabilities. However, it may be possible to construct the effective hydrodynamic equation explicitly with our method. A straightforward generalization of the 1D analysis points to some important differences between the 1D and 2D systems: The configurations of the cells can no longer be described by a scalar density field, since the local orientational degrees of freedom must be specified in 2D. Also, the local cell arrangements are determined by the local curvature tensor, leading to more complex elastic responses. The minimal effective equation one can construct in this case is the analog of Eqs. (4.13) and (4.14), with $\pi(x,t)$ generalized to a complex field, whose phase describes the orientational degrees of freedom. Recall that symmetry is valuable in constructing the effective hydrodynamic equation only if *all* of the slow degrees of freedom have been identified. The existence (and possible relevance) of the orientational degrees of freedom in the 2D KS system therefore makes the correspondence to KPZ dynamics less transparent. A study of the 2D KS dynamics will be reported elsewhere.

On a broader scope, it is important to see to what extent the approach presented can be applied to characterizing and understanding different types of STC. It seems most suited to describing phase turbulence where space-time dislocations or defects are interspersed in a background pattern that is otherwise periodic. Our analysis suggests that effective hydrodynamic equations may be obtained by simply breaking wavenumber conservation and including a noise source in the appropriate slow mode equations describing the long wavelength distortions of the *periodic* patterns (the analog of Equations (3.11) and (3.12)). On the other hand, due to the weak spatial-coupling assumption used, this approach probably may not be helpful in understanding STC in coupled chaotic maps which rely crucially on the spatial coupling among neighboring maps. For such problems, the method recently introduced by Hansel and Sompolinsky [31] may be more appropriate. Through the exploration of these issues, one can hopefully gain more insight into the classification and characterization of various types of spatiotemporal chaos.

## VIII. ACKNOWLEDGMENT

During the course of this work, we benefitted from discussions with Bruce Boghosian, Alastair Rucklidge, and Haim Sompolinsky. TH is grateful to the hospitality of the Niels Bohr Institute and the Newton Institute of Mathematical Sciences, where part of this work was completed. CC was supported by US Department of Energy Grant No. DE-FG02-86ER40302. TH was supported by US Department of Energy Grant No. DE-FG02-90ER40542 and an A. P. Sloan Fellowship.

# APPENDIX A: CONSTRUCTION OF THE LANGEVIN EQUATION

We review the formal construction of a stochastic equation for a time series $w(t)$ subject to a time-dependent perturbation $\epsilon(t)$ given the leading moments of the distribution $P[w;\epsilon]$. It is convenient to consider the Fourier transform of $P$,

$$\widehat{P}[\widehat{w};\epsilon] = \int \mathcal{D}w \ P[w;\epsilon] \ e^{-i \int dt \ \widehat{w}w}, \tag{A1}$$

which generates moments of $P$ through the Green functions

$$G_{mn}(t_1, \cdots, t_m, t_{m+1}, \cdots, t_{m+n}) = (i)^n \frac{\delta^{m+n} \log \widehat{P}[\widehat{w};\epsilon]}{\delta\epsilon(t_1) \cdots \delta\epsilon(t_m) \delta\widehat{w}(t_{m+1}) \cdots \delta\widehat{w}(t_{m+n})}. \tag{A2}$$

For example,

$$G_{01}(t) = \langle w(t) \rangle \equiv v, \tag{A3}$$

$$G_{02}(t_1, t_2) = \langle (w(t_1) - v)(w(t_2) - v) \rangle \tag{A4}$$

are just the mean and variance of the distribution $P[w;0]$, and

$$G_{11}(t_1, t_2) = \frac{\delta}{\delta\epsilon(t_1)} \langle w(t_2) \rangle \tag{A5}$$

is the linear response function. Writing $\widehat{P} = e^{\widehat{S}}$, we see that $G_{mn}$ are just the coefficients of the Taylor expansion of $\widehat{S}$:

$$\widehat{S}[\widehat{w},\epsilon] = \sum_{m,n=0}^{\infty} \int dt_1 \cdots dt_{m+n} \frac{(-i)^n}{m!n!} G_{mn}(t_1, \cdots, t_{m+n})$$
$$\epsilon(t_1) \cdots \epsilon(t_m) \widehat{w}(t_{m+1}) \cdots \widehat{w}(t_{m+n}). \tag{A6}$$

The "dynamic action" $\widehat{S}$ can now be used to construct the effective equation of motion. Suppose the leading moments of $P[w;\epsilon]$ [Eqs. (A3), (A4) and (A5)] exist, and the higher order moments are negligible. In this case, the action becomes

$$\widehat{S}[\widehat{w},\epsilon] = -\frac{1}{2} \int dt dt' \ G_{02}(t,t') \widehat{w}(t)\widehat{w}(t') - iv \int dt \ \widehat{w}(t) - i \int dt dt' \ G_{11}(t,t')\epsilon(t)\widehat{w}(t'). \tag{A7}$$

Rewriting the first term of Eq. (A7) by another Fourier transform, we have

$$\hat{P}[\hat{w};\epsilon] = \int \mathcal{D}\eta \, \exp\left\{-i\int dt \, [v+\eta(t)]\hat{w}(t) - i\int dt dt' \, G_{11}(t,t')\epsilon(t)\hat{w}(t')\right\} \times P[\eta], \quad (A8)$$

where

$$P[\eta] \propto \exp\left\{-\frac{1}{2}\int dt dt' \, G_{02}^{-1}(t,t')\eta(t)\eta(t')\right\}, \quad (A9)$$

with $G_{02}^{-1}(t,t')$ denoting the inverse of $G_{02}(t,t')$. Inverse Fourier transforming Eq. (A8) leads to the result

$$\begin{aligned} P[w;\epsilon] &= \int \mathcal{D}\hat{w}\hat{P}[\hat{w},\epsilon]\, e^{i\int dt \, \hat{w}(t)w(t)} \\ &= \int \mathcal{D}\eta \prod_t \delta\left[w(t) - v - \eta(t) - \int dt' G_{11}(t',t)\epsilon(t')\right] \times P[\eta]. \end{aligned} \quad (A10)$$

Eq. (A10) clearly describes a time series which satisfies the "equation of motion"

$$\int dt' \, G_{11}^{-1}(t,t')\left[w(t') - v\right] = \xi(t) + \epsilon(t), \quad (A11)$$

where

$$\xi(t) = \int dt' G_{11}^{-1}(t,t')\eta(t') \quad (A12)$$

plays the role of a Langevin noise. By the distribution (A9), we see that $\xi$ is Gaussian distributed, with the variance

$$\langle \xi(t)\xi(t')\rangle \equiv \Delta(t-t') = \int d\tau d\tau' G_{11}^{-1}(t,\tau)G_{11}^{-1}(t'-\tau')G_{02}(\tau-\tau'). \quad (A13)$$

In the presence of higher moments of $P[w;\epsilon]$, it is straightforward to verify that $P[\eta] = P[v+\eta;0]$. In addition, the equation of motion (A11) also acquires nonlinear terms. However for the KS equation, the nonlinear terms as well as the higher moments of $P[\eta]$ are irrelevant in the hydrodynamic limit and are not pursued here.

In Sec. V, we find the linear response of the average drift rate $\langle w(t)\rangle$ of a small system to a curvature perturbation $\epsilon(t) = \partial_x^2 h_< \theta(t)$ to be well described by the exponential form

$$\langle w(t)\rangle = v + \frac{\beta}{\alpha}\left(1 - e^{-\alpha t}\right)\partial_x^2 h_<. \quad (A14)$$

Since $\langle w(t)\rangle = v + \int dt' G_{11}(t',t)\epsilon(t')$ according Eq. (A10), we find

$$G_{11}(t', t) = \beta e^{-\alpha(t-t')}\theta(t - t'). \tag{A15}$$

Inverting Eq. (A15) yields the operator

$$G_{11}^{-1}(t, t') = \delta(t - t')(\partial_t + \alpha)/\beta. \tag{A16}$$

Finally, applying Eq. (A16) in Eq. (A11), we obtain the equation of motion

$$\partial_t w = -\alpha(w(t) - v) + \beta\partial_x^2 h_< + \xi(t), \tag{A17}$$

with

$$\langle \xi(t)\xi(t') \rangle \equiv \Delta(t - t') = (\alpha^2 - \partial_t^2)G_{02}(t - t') \tag{A18}$$

to leading order in $w(t) - v$ and $\xi$.

# REFERENCES

\* Present Address: NeuroMuscular Research Center, Boston University, 44 Cummington St., Boston MA 02215.

\*\* Present Address: Department of Physics, SUNY at Stony Brook, Stony Brook, NY 11794-3800.

[16] U. Frisch, Z.S. She, and O. Thual, J. Fluid Mech. **168**, 221 (1986).

[17] V. Yakhot, Phys. Rev. A **24**, 642 (1981).

[18] D. Forster, D. R. Nelson, and M. J. Stephen, Phys. Rev. A **16**, 732 (1977).

[19] M. Kardar, G. Parisi, and Y.-C. Zhang, Phys. Rev. Lett. **56**, 889 (1986).

[20] E. Medina, T. Hwa, M. Kardar and Y.-C. Zhang, Phys. Rev. A **39**, 3053 (1989).

[21] A.M. Rucklidge, J. Fluid Mech. **237**, 209 (1992);

[22] P.H. Coullet and E.A. Spiegel, SIAM J. Appl. Math **43**, 776 (1983).

[23] This Ansatz assumes *instantaneous* adjustment of cell shape to the one corresponding to its size.

[24] To obtain Eq. (3.8), we need to use $l(x,t)$ instead of $\ell$ as the coarse-graining scale. This results in a slight modification of the slow mode equation (3.4). The difference is however not important in the hydrodynamic limit.

[25] Note that $w_<(x,t)$ only depends on $l(x,t)$, not $\ell$. Also, it does not 'know' about $k_0$, which is the preferred wavenumber of the *linear* stability analysis.

[26] H. Chaté and P. Manneville, Phys. Rev. Lett. **58**, 112 (1987); H. Chaté, Ph. D thesis, Universite Pierre et Marie Curie, (1989).

[27] R. L. Chuoke, P. Van Meurs and C. Van der Pol, Tr. AIME **216**, 188 (1959).

[28] F. Hayot, C. Jayaprakash, and Ch. Josserand, Phys. Rev. E **47**, 911 (1993).

[29] C. Jayaprakash, F. Hayot, and R. Pandit, Phys. Rev. Lett. **71**, 12 (1993).

[30] I. Procaccia, M.H. Jensen, V.S. L'vov, K. Sneppen, and R. Zeitak, Phys. Rev. A **46**, 3220 (1992); V. S. L'vov and I. Procaccia, Phys. Rev. Lett. **69**, 3543 (1992).

[31] D. Hansel and H. Sompolinsky, Phys. Rev. Lett. **71**, 2710 (1993).

FIGURES

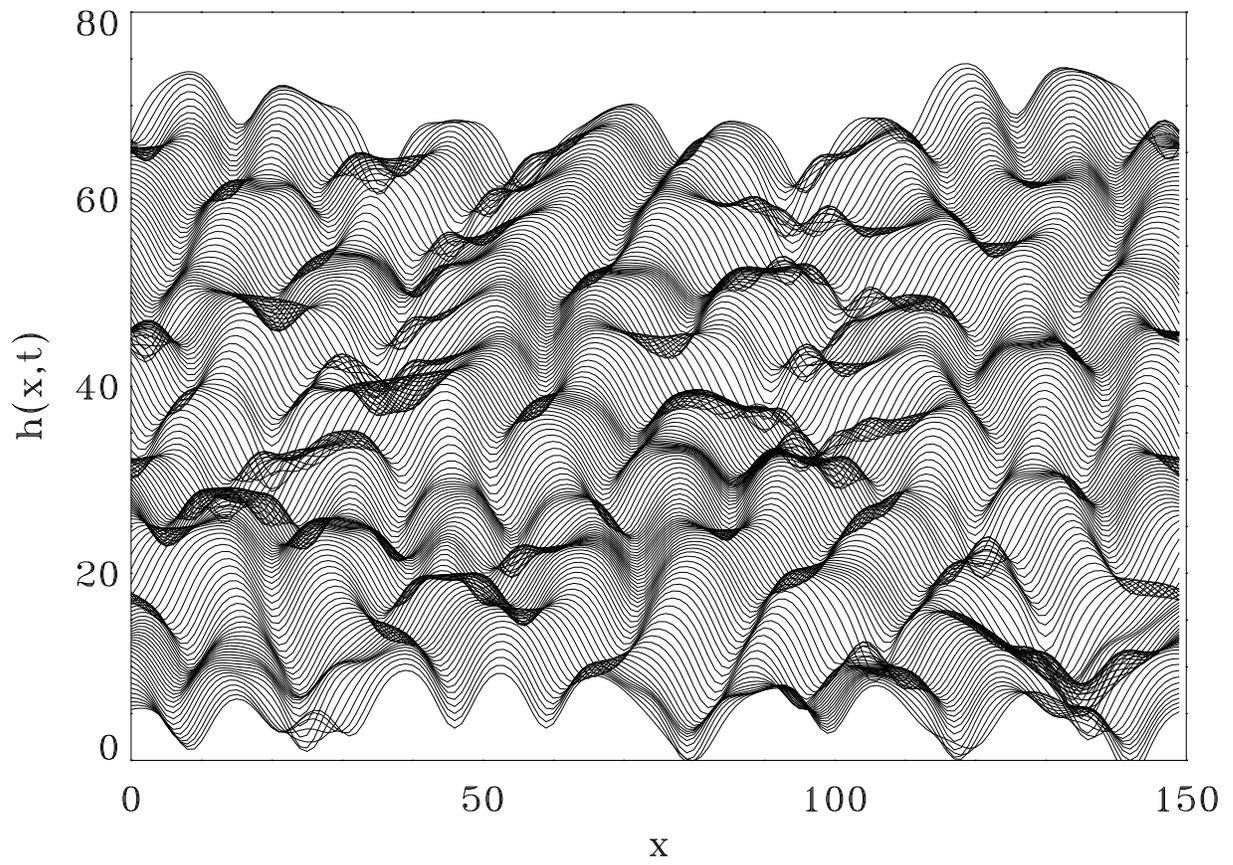

FIG. 1. Spatiotemporal chaos produced by the KS dynamics.

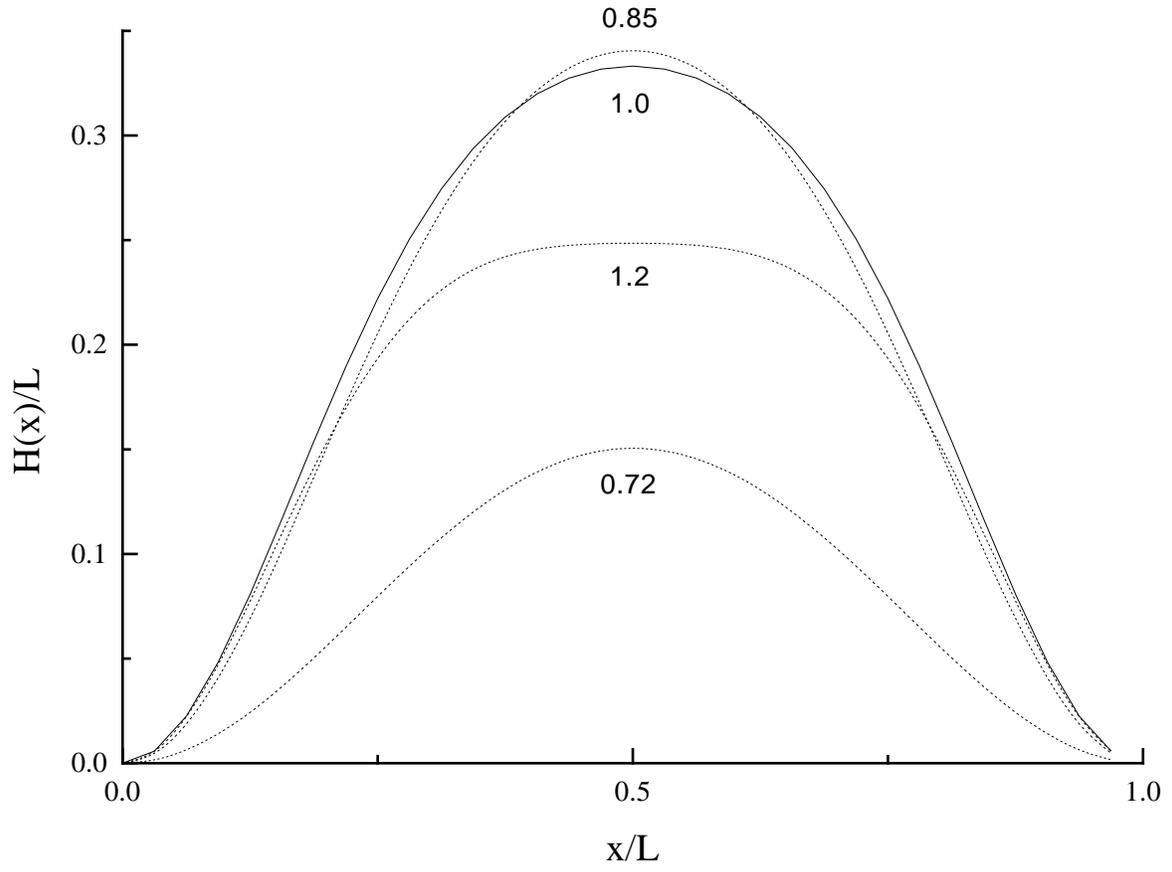

FIG. 2. Shape of the cellular solutions for $L/l_0 = 0.72, 0.85, 1.0$ (solid line), and $1.2$. Both the width and height of the cells are normalized by $L$.

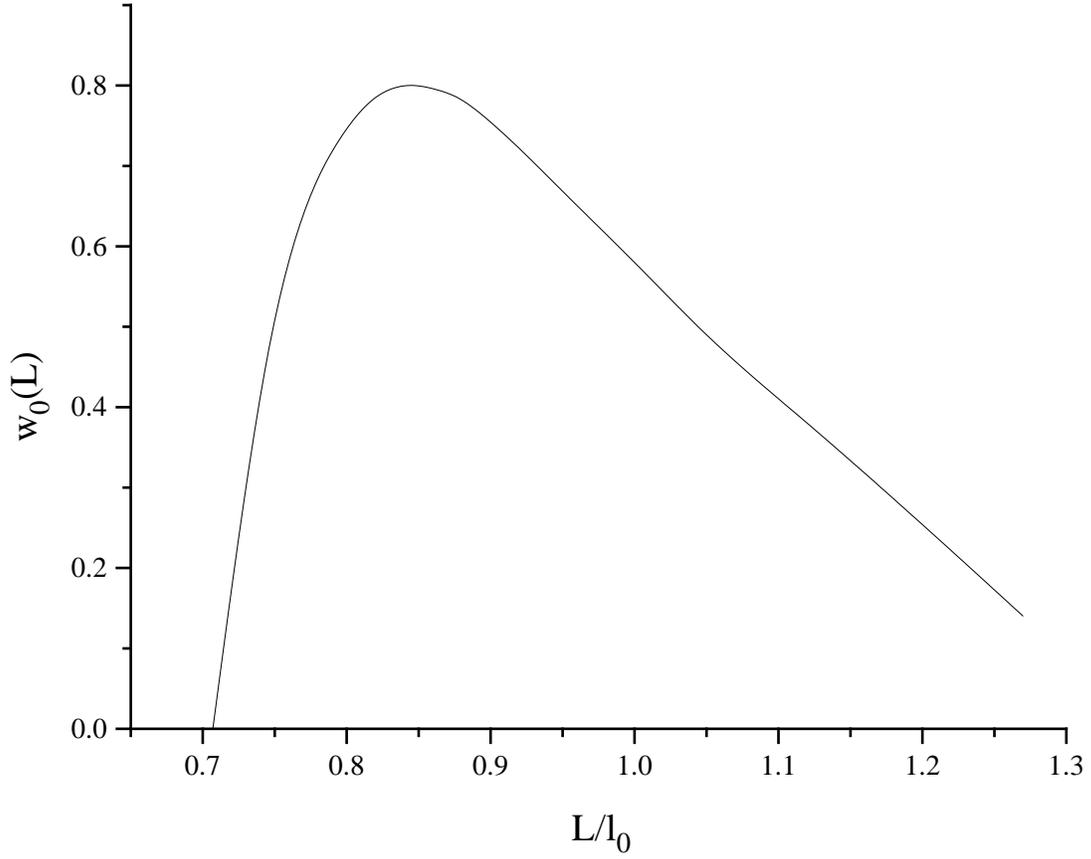

FIG. 3. Size-dependence of the single-cell drift rate $w_0(L)$. $l_0 = 8.89$ is the most unstable wavelength according to the linear stability analysis. $l_1 = 0.707 l_0$ is the length where the nontrivial uni-cellular solution first appears. $l_2 = 1.27 l_0$ is the maximum size of a stable uni-cellular solution. $w_0(L)$ has a maximum at $L = l^* = 0.85 l_0$.

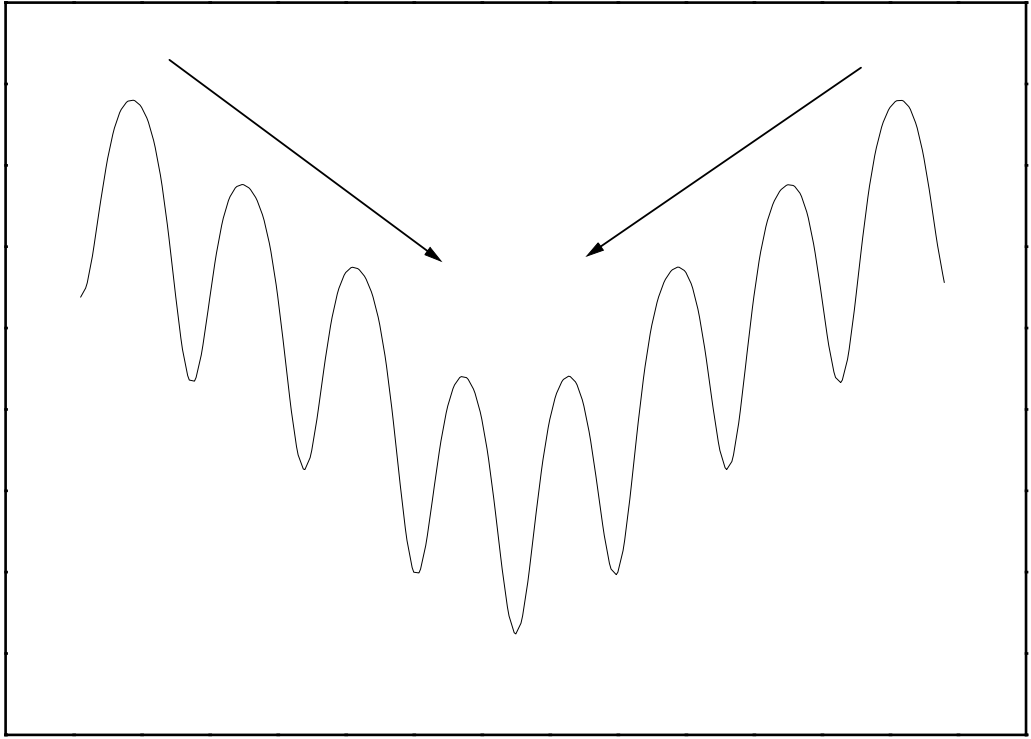

FIG. 4. Galilean invariance requires the cells to advect in the downhill direction, leading to accumulation of cells in valleys and depletion of cells on hilltops.

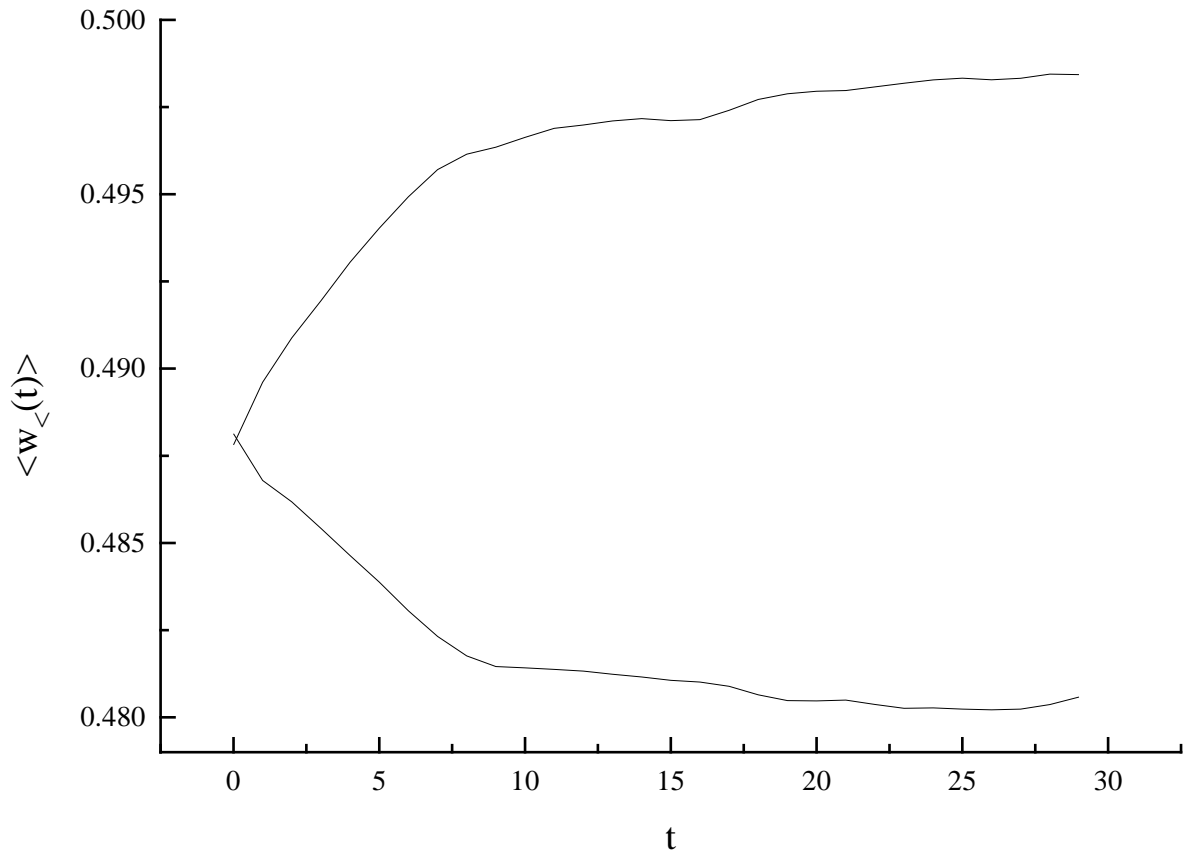

FIG. 5. Time evolution of the average drift rate $\langle w_<(t) \rangle$ after a small curvature forcing ($c' = 0.002$) is turned on at $t = 0$ after the transients have died away. The data is taken for a 128-site system and averaged over $10^5$ configurations of random initial conditions. The upper (lower) branch corresponds to the drift rate measured from the left (right) half of the system.

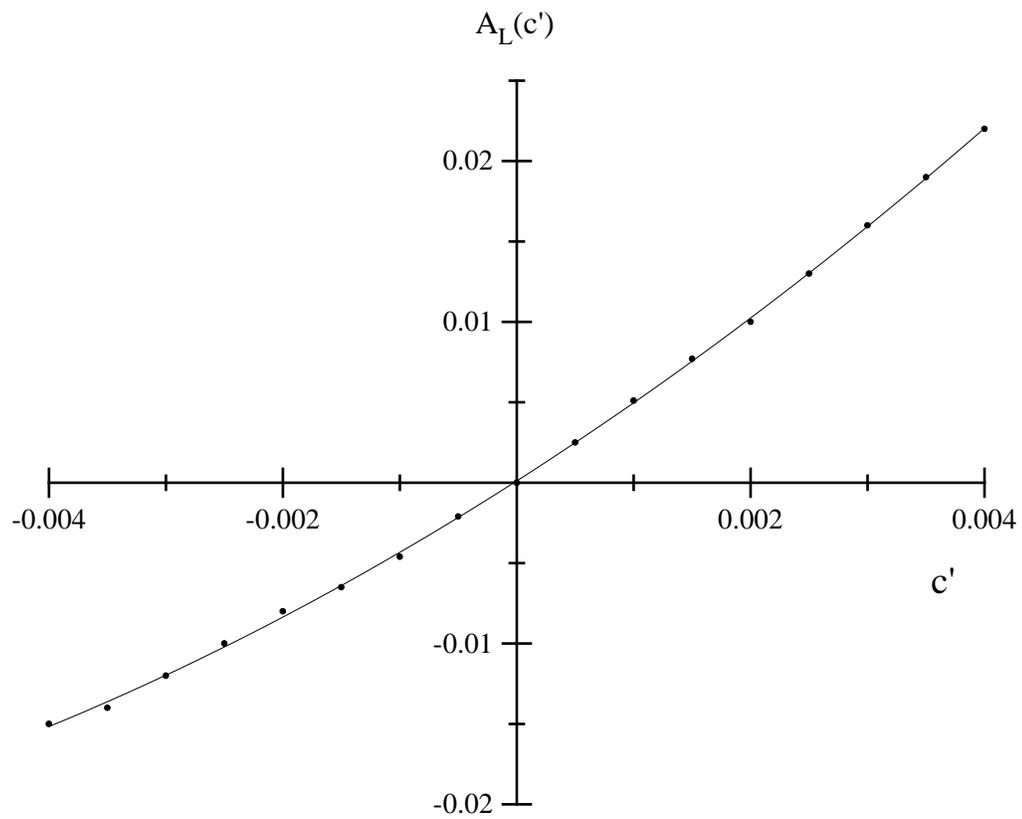

FIG. 6. The saturated response $A_L(c')$ of an $L = 64$ system for various imposed curvature $c'$. The solid line is a least-square fit (see the text).

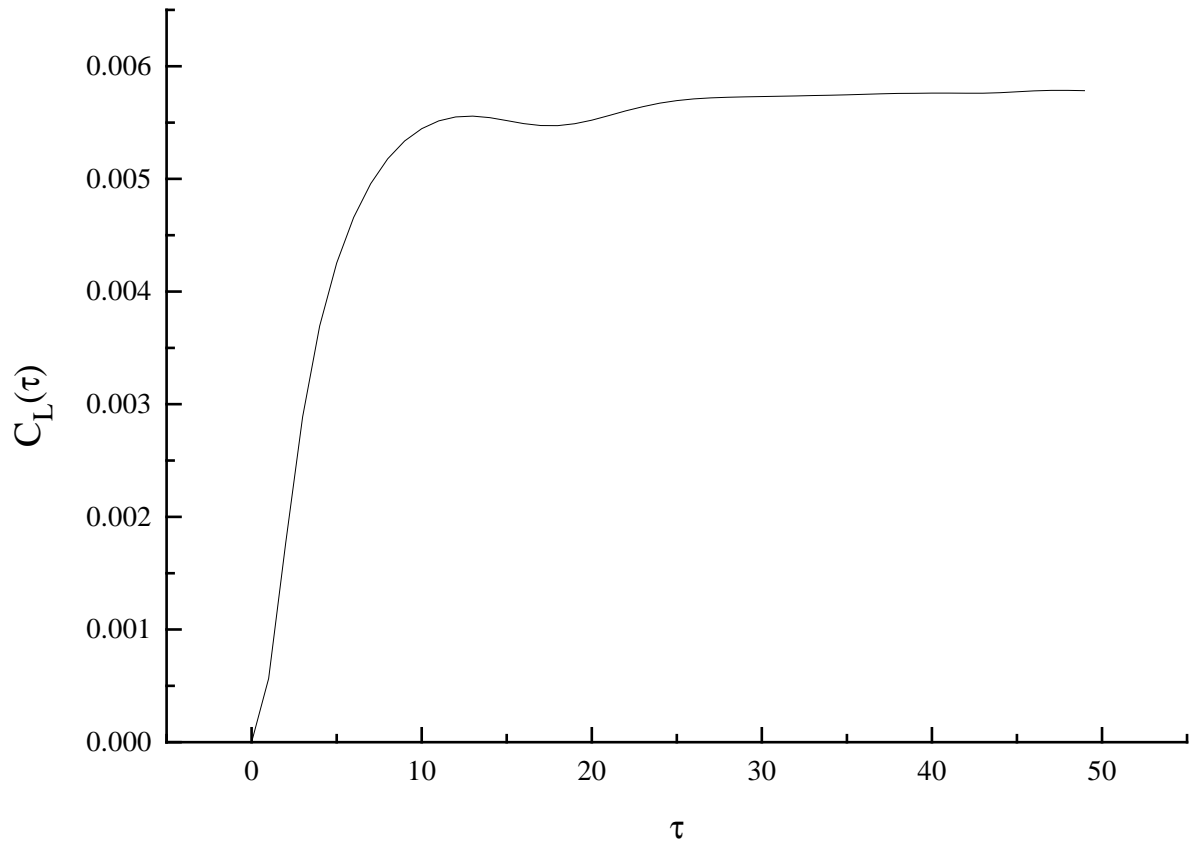

FIG. 7. The auto-correlation function $C_L(\tau)$ of the drift rate of an $L = 100$ system. The data shown is averaged over $10^4$ configurations of random initial conditions.

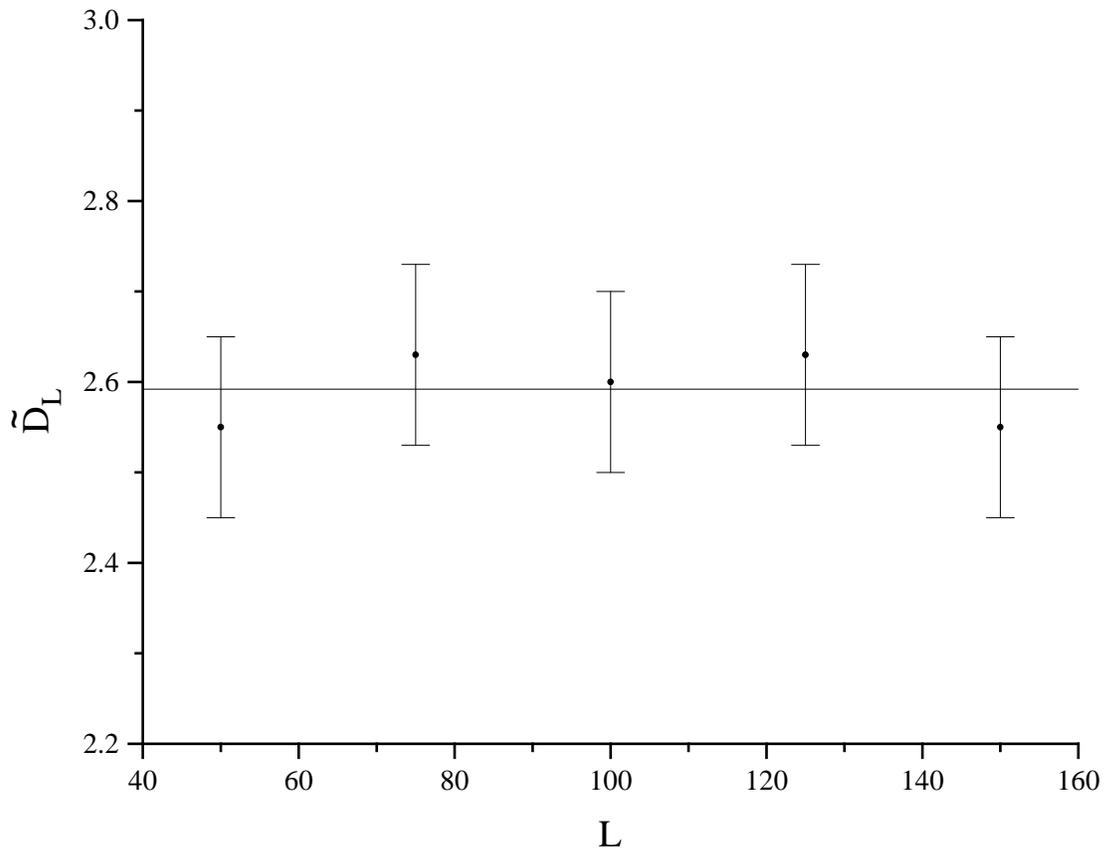

FIG. 8. The noise amplitude $\widetilde{D}_L$ as a function of $L$. The solid line shows no $L$ dependence.

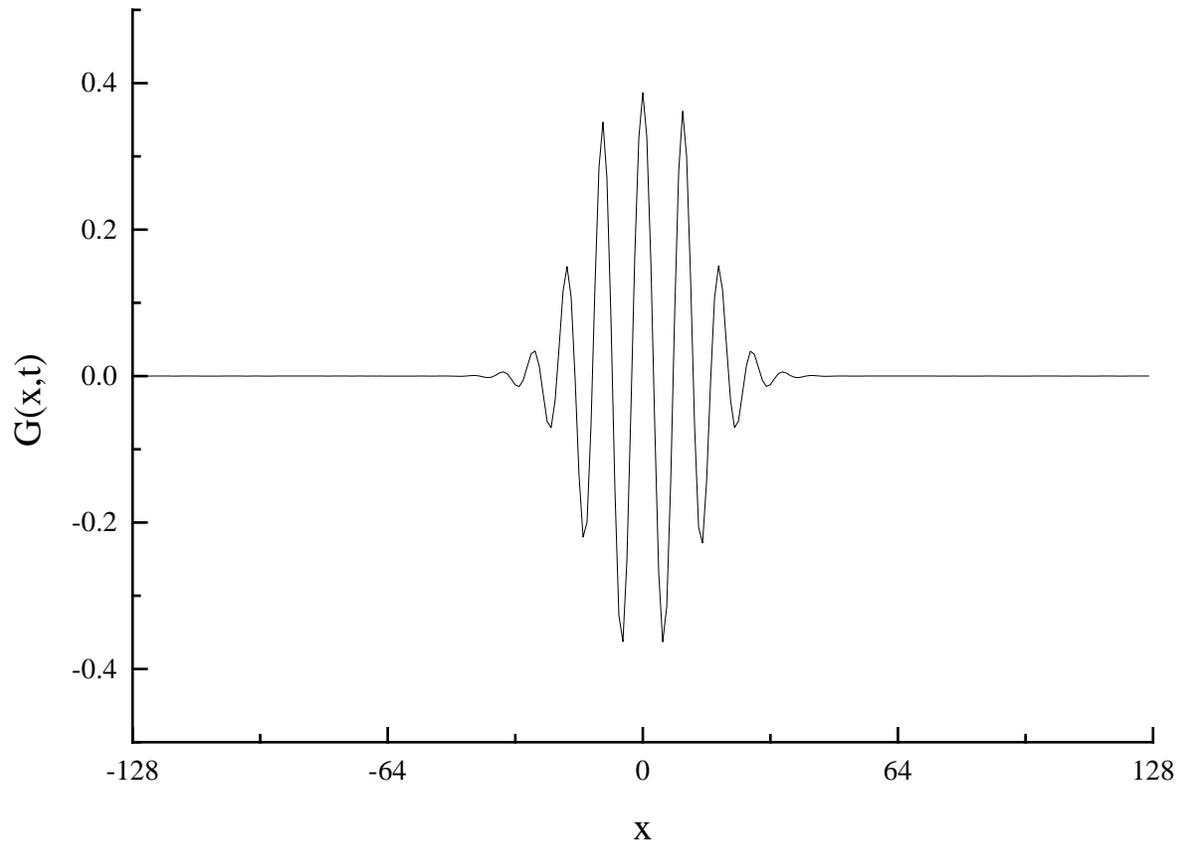

FIG. 9. The response of the KS equation to an infinitesimal perturbation of the initial conditions at $x = 0$. The result is averaged over 51,200 configurations of random initial conditions. The measurement was performed on a CM2 connection machine.

43